# Rahmani Sort: A Novel Variant of Insertion Sort Algorithm with O(nlogn) Complexity

## Mohammad Khalid Imam Rahmani[1, *]

[1]Department of Computer Science, College of Computing and Informatics, Saudi Electronic University, Riyadh 11673, Saudi Arabia

*Corresponding Author: Mohammad Khalid Imam Rahmani. Email: m.rahmani@seu.edu.sa

**Abstract**: Due to the abundance of large number of data repositories with ever-growing volume of online and offline data which are being maintained by enterprise houses, research institutions, medical & healthcare organizations, finding a key is a time-consuming task. For taking a strategic decision, the managers of such organizations analyse the related data for selecting the optimal option among the available choices. For this various decision support systems are available which implement Data Mining and Data Warehousing techniques for diving into the sea of data for getting the useful patterns of knowledge (pearls). Classification, regression, clustering and many other algorithms are used for enhancing the precision and accuracy of the decision process. So, there is scope of increasing the response time of decision process specially in mission-critical operations. If data are ordered with suitable and efficient sorting operation, response time of the decision process can be minimized. Insertion sort is much suitable for such applications due to its simple and straight logic along with its dynamic nature suitable to list implementation. But it is slower than merge sort and quick sort. The main reasons that why this is slow: firstly, sequential search is used to find the actual position of the next key element into the sorted left sub array and secondly, shifting of elements is required by one position towards the right for accommodating the newly inserted element. Therefore, I propose a new algorithm by using a novel technique of binary search mechanism for finding the sorted location of the next key item into the previously sorted left sub array much quicker than conventional insertion sort algorithm. Performance measurement in terms of actual running time of the new algorithm has been compared with those of other conventional sorting algorithms apart from the insertion sort. The results obtained on various sample data show that the new algorithm is better in performance than the conventional insertion sort and merge sort algorithms.

## 1 Introduction

Many sorting algorithms have emerged since the beginning of the philosophy of computer programming [1- 4]. But still many improvements have been introduced in sorting algorithms during the last decade [5-14]. Sorting is the process of arranging the elements in some ordered sequence which can be either in ascending, descending or lexicographic order [2-3] [15-16]. It is estimated that more than 25% of all computing efforts or CPU cycles are spent on sorting of the data [17]. As a matter of fact, much research on the topic of sorting has been done. But there is not a single sorting technique which can be considered the best among the rest. Bubble sort, selection sort and exchange sort are having asymptotic complexity of O($n^2$). Therefore, they are applicable for small input size. Insertion sort is having asymptotic complexities of $\Omega(n)$ and O($n^2$). Therefore, it is applicable for medium input size. Quick sort is having asymptotic complexities of $\Theta(n\log n)$ and O($n^2$) but it is considered the fastest algorithm in average case. Therefore, it is applicable for large input size. Merge sort and heap sort are having asymptotic complexity of $\Theta(n\log n)$. Therefore, they are applied for large input size [2-4] [15] [18]. Selection of a faster sorting algorithm is one of important factors for improving the performance of underlying data processing applications. For selecting a faster sorting algorithm, generally their asymptotic complexity analyses are performed. With the help of asymptotic complexity analysis, we can select as per the priority of $\Omega(n)$, $\Theta(n\log n)$ and O($n^2$) respectively. But they cannot be distinguished if they have the same asymptotic complexity. Therefore, empirical testing of sorting algorithms is required in such situations [19-21].

These sorting algorithms are comparison-based algorithms and hence they cannot be faster than O($n\log n$) [3] [1], where O is the notation called Big oh, used for measuring the order of growth of any algorithm in worst case scenario for a data size of $n$. There are a few algorithms claiming to run in linear time for special cases of input data. So, there is a scope and requirement for working out of a new sorting algorithm which can be easy to implement on all modern devices for all permissible input data and may also beat the lower bound O($n\log n$) of sorting algorithms for some cases. This work is an effort towards that direction.

In this paper, a new enhanced sorting algorithm has been designed which shows more efficiency than the classical insertion sort and other sorting algorithms like bubble sort, quick sort and merge sort.

The technique used for designing the proposed algorithm has been inspired from the technique used in binary search; that is for quickly finding the location of the next element to be placed in the sorted sub array. The sequential search technique used to find the location of the next element in the sorted sub array takes much more time due to the fact that it has to compare the next element with every element of the sorted sub array starting from its rightmost element until its correct location is found.

The entire paper is organised in the following manner. The introductory concept of the subject matter has briefly been explained in this section (section I). In section II, the step-by-step method of the insertion sort is explained after some background work related to the general sorting technique. The other sorting algorithms like merge sort and quick sort are explained in section III. The new sorting algorithm is introduced and discussed in section IV. The Java based performance measurement Framework and performance analysis of the new algorithm is done in section V. Results and comparison of performance of various sorting algorithms have been discussed in tabular and graphical forms in section VI. Finally, the conclusions have been drawn and future scope of the research is mentioned in the section VII.

## 2 Preliminaries

Sorting is a process of arranging the available data items into an ordered sequence [2-3]. The process of sorting is applied to a collection of items prior to any such operation which may consume more time and/or space if applied without prior sorting [3-4]. An efficient sorting mechanism is important to optimizing the design of other algorithms that require sorted data items for further processing. A sorting algorithm is a set of steps arranged in a particular sequence that puts the available data items into a certain order. The known ordered sequences have been increasing order, decreasing order, non-increasing order, non-decreasing order and lexicographic order.

Let $r_1, r_2, r_3, ..., r_n$, be n number of input data items. For the output data items to be in sorted order, one of the following conditions must be satisfied:

***Increasing order:*** For all i, $1 \leq i \leq n$, $r_i < r_i+1$.

***Decreasing order:*** For all i, $1 \leq i \leq n$, $r_i > r_i+1$.

***Non-decreasing order:*** For all i, $1 \leq i \leq n$, $r_i \leq r_i + 1$.

***Non-increasing order:*** For all i, $1 \leq i \leq n$, $r_i \geq r_i + 1$.

***Lexicographic order:*** Words in a dictionary of the English language are arranged in this order.

## 2.1 Definition of Sorting

A sorting technique can be defined based on partial order relation. The definition of partial order is given as below:

***Definition 1.*** Let R be a relation on a set S. For a, b, c ∈ S, if R is:

*a) Reflexive*, i.e., aRa for every a ∈ S;

b) *Transitive*, i.e., aRb ∧ bRc ⇒ aRc; and

*c) Antisymmetric*, i.e., aRb ∧ bRa ⇒ a = b,

then, R is a partial order on the set S.

Sorting is generally defined as an arrangement of a list of randomly input data by their key or themselves into a partial order R, where R implies ≤ in particular.

***Definition 2.*** For N elements a(1), a(2), ..., a(n) ∈ S, sorting is rearrangement of the elements in order to obtain a partial order $a(s_i)$ R $a(s_i+1)$ for ∀ $s_i$, $1 \leq s_i < n$. Generally, R is defined as ≤ in sorting, so that the partial order is: $a(s_1) \leq a(s_2) \leq, ..., \leq a(s_i) \leq, ..., \leq a(s_n)$. For example, suppose the given input sequence is ⟨13, 4, 1, 45, 30, 8, 10, 7, 5⟩. A valid sorting algorithm will return as output the sequence ⟨1, 4, 5, 7, 8, 10, 13, 30, 45⟩.

## 2.2 Importance of Sorting in Computation

There are two direct applications of sorting: first as an aid for searching and second as a tool to match entries in files. Broad areas of application of sorting fall in the solution of many other more complex problems, from database systems, networking, MIS, operations research and optimization problems. Sorting algorithm is one of the most fundamental techniques in computer science because of the following reasons. First, it is the basis of many other algorithms such as searching, pattern matching, information retrieval, knowledge-based systems, digital filters, database systems, data statistics and processing, data warehousing, and data communications [1]. Second, it plays an important role in the teaching of design and analysis of algorithms, programming methodology, data structures and programming. Furthermore, it is a very challenging problem which has been widely and thoroughly studied [19-24]; the performance is dramatically improved [25-30] and considered the lower-bound of complexity has been reached [19-20] [29-30].

### 2.2.1 Applications of Sorting

There are many applications of sorting. Once a list is sorted many questions about the list can be answered easily. We can efficiently find an element in a sorted list using Binary Search. Binary search requires only O(log$n$) operations in finding an element. We can also determine in O($n$) if a sorted list has duplicates. We can construct a frequency distribution of the list if the list is sorted, or find the median and mode of the list in O(*1*) and O($n$) respectively. We can find the kth largest element in a list in O(*1*) time. In Big data sorting is one of the basic operators.

## 2.3 Classification of Sorting Algorithms

Sorting algorithms may be organized on the basis of the classification of the underlying technique or strategy used to rearrange the data items in the given sequence. If the items to be sorted all fit into the computer's internal memory, then it is known as an internal sorting algorithm. Due to the growing power of computers, external storage devices become less frequent in sorting. If all the items cannot be stored in the internal memory at one time, different techniques have to be used. The underlying idea is to sort as many items as the internal memory can handle at a time and then merge the results into external storage devices. Sorting algorithms are also classified by their computational complexity and ease of implementation. For a typical sorting algorithm ideal behavior is O($n$), good behavior is O($n\log n$) and bad behavior is O($n^2$). The lower bound of time complexity of sorting algorithms, which only use key comparison operation, is O($n\log n$). Comparison based sorting algorithms rearrange the input data items by comparing the key values of the adjacent items or that of one item with that of another item. Some sorting algorithms are in-place, such that only O($1$) or O($\log n$) memory is needed in addition to the items being sorted, while others require auxiliary memory locations for data to be temporarily stored. Some sorting algorithms are recursive and some are non-recursive, whereas some may be implemented in both ways.

## 2.4 Performance Measurement/Analysis of Sorting Algorithms

The performance of an algorithm is gauged in terms of its measure of efficiency, which is measured on the basis of its time complexity and space complexity (collectively known as computational complexity). Time complexity tends to be more important as availability of memory nowadays is not a big deal. The efficiency of an algorithm is always stated as a function of input data size i.e., as T($n$) or S($n$), where $n$ is the input data size.

All spreadsheet programs and database applications use some sorting code. Because of the importance of sorting in these applications, dozens of sorting algorithms have been developed over the decades with varying complexity. Slow sorting methods such as bubble sort, insertion sort, and selection sort have a theoretical complexity of O($n^2$). Even though these algorithms are very slow for sorting large arrays, the algorithm is simple, so they are not useless. If an application only needs to sort small arrays, then it is satisfactory to use one of the simple slow sorting algorithms as opposed to a faster, but more complicated sorting algorithm. For these applications, the increase in coding time and probability of coding mistake in using the faster sorting algorithm is not worth the speedup in execution time. Of course, if an application needs a faster sorting algorithm, there are certainly many ones available, including quick sort, merge sort, and heap sort. These algorithms have a theoretical complexity of O($n\log n$). They are much faster than the O($n\log n$) algorithms and can sort large arrays in a reasonable amount of time. However, the cost of these faster sorting methods is that the algorithm is much more complex and is harder to correctly code. But the result of the more complex algorithm is an efficient sorting method capable of being used to sort very large arrays.

In addition to varying complexity, sorting algorithms also fall into two basic categories: comparison based and non-comparison based. A comparison-based algorithm orders a sorting array by weighing the value of one element against the value of other elements. Algorithms such as quick sort, merge sort, heap sort, bubble sort, and insertion sort are comparison based. A non-comparison-based sorting algorithm sorts an array without consideration of pair wise data elements. Radix sort is a non-comparison-based algorithm that treats the sorting elements as numbers represented in a base-M number system, and then works with each digit of M.

Mathematicians and computer scientists have been researching and analyzing their performance for many years. The performance analysis was mostly based on the theory behind the algorithm. D.E. Knuth's book, "The Art of Computer Programming, Volume III - Sorting and Searching" is considered the bible for sorting algorithms because of his detailed analysis of many sorting algorithms. However, even though Knuth gives a complete analysis of the different algorithms, they are all based on non-cached computer architectures. All of the analysis is based on the theoretical complexity of the algorithm. But as the cached computer architecture becomes common today, it is necessary to analyze how a cached memory affects the performance of these sorting algorithms. It is not to say that the theoretical analysis is useless. They are still useful because those are the fundamental analysis that is needed in analyzing any kind of algorithms. Even though there is an abundance of previous research on the performance of sorting algorithms, most of the research does not analyze how the sorting algorithms exploit cache. As all of today's computers contain cached memory architecture, this is an area that is definitely lacking in research. In addition, as the increase in memory access time is larger than the increase in processor cycle time, the cache performance of an algorithm will have an increasingly larger impact on the overall performance.

## 2.5 Growth of Functions

The order of growth of the execution time of an algorithm gives a simple characterization of the algorithm's efficiency and also allows us to compare the relative performance of the alternative algorithms for solving the same problem. Although we may sometimes require to determine the CPU execution time of an algorithm, as it has been done for some sorting algorithms in this paper, the extra precision is not usually worth the effort of computing it. For large enough inputs, the multiplicative constants and lower-order terms of an exact running time are dominated by the effects of the input size itself. When we look at input sizes large enough to make only the order of growth of the running time relevant, then we are analyzing the asymptotic efficiency of algorithms. That is, we are concerned with how the running time of an algorithm increases with the size of the input in the limit, as the size of the input increases without bound. Usually, an algorithm that is asymptotically more efficient will be the best choice for all but very small inputs.

## 2.6 Cases for Analysis

In order to compare different algorithms for the problem of sorting, analysis of the algorithms, as usual, is broken into three different cases so that some way of providing a better estimate of the resources required can be formulated.

### 2.6.1 Best Case

In this case, the minimum numbers of computational steps are taken by the algorithm on any input size. This is a trivial case having less practical importance but gives the estimate of the lowest resource requirement.

### 2.6.2 Average Case

In this case, the average numbers of computational steps are taken by the algorithms on any input size. This is the most complex analysis often based on the probability theory. Proper formulation of this case depends on the statistical distribution of the input data. But in actual sense this is very complex analysis for the case of sorting algorithms. Provide some deal from the permutation/combination of input data.

### 2.6.3 Worst Case

In this case, the maximum numbers of computational steps are taken by the algorithms on any input size. This case is most often analyzed because maximum time complexity of algorithm occurs in this case. The upper bound of the resources required by any algorithm is one of the important measures to be considered for selecting an efficient algorithm for solving a given problem.

## 2.7 Comparing Performance of Sorting Algorithms

In the **Tab. 1**, comparisons of the performance of above algorithms that operate on arrays are being mentioned. These numbers are only of small significance giving order of measurement; they can vary from one machine architecture to other machine architecture. Quick sort and merge sort are algorithms where the recursive procedure is switched to Selection sort once the size of the array falls to 16 and below. There is usually a best sorting algorithm under the given circumstances and it is up to the developer to pick the appropriate one.

**Table 1:** Sorting algorithms, relevant data structures and time complexities

| Algorithm | Method | Data Structure | Best Case | Worst Case | Average Case | Stable |
|---|---|---|---|---|---|---|
| Bubble sort | Exchange | Arrays | $\Omega(n^2)$ | $O(n^2)$ | $\Theta(n^2)$ | Yes |
| Selection sort | Selection | Arrays | $\Omega(n^2)$ | $O(n^2)$ | $\Theta(n^2)$ | No |
| Insertion sort | Insertion | Arrays/Lists | $\Omega(n)$ | $O(n^2)$ | $\Theta(n^2)$ | Yes |
| Quick sort | Partitioning | Arrays | $\Omega(n\log n)$ | $O(n^2)$ | $\Theta(n\log n)$ | No |
| Merge sort | Merge | Arrays/Lists | $\Omega(n\log n)$ | $O(n\log n)$ | $\Theta(n\log n)$ | Yes |
| **Rahmani sort** | Search/Insertion | Arrays/Lists | $\Omega(n)$ | $O(n^2)$ | $\Theta(n^2)$ | Yes |

## 3 The Proposed Algorithm

The concept of the proposed algorithm, its formal steps and their explanations are being described in this section. The understanding of all these concepts is very necessary for proper analysis of the algorithms.

### 3.1 The Concept

In the classical insertion sort, the original array is safely assumed to have two logical subarrays, the left sorted subarray and the right probably unsorted subarray. At the start of the algorithm, only the first element of the array is kept in the left subarray and the rest of all elements are kept in the right subarray. Now, the first element from the right subarray is placed into the proper position of the sorted left subarray. But while finding the proper position of the element in the left subarray, a simple sequential search technique is used which has a time complexity of $O(n)$. Even this linear time complexity for searching the proper location of the element to be inserted may be quite considerable. That is why insertion sort is not considered a suitable sorting algorithm for sorting of large number of elements. So, by implementing some novel way of expediting the search technique for the proper location of the element adopted in insertion sort algorithm, the performance of sorting can be improved.

The proposed new sorting algorithm, hereafter called Rahmani sort algorithm, is based on the novel concept of inserting the elements of the unsorted right sub array starting with its first element into the proper position of the sorted right sub array. The classical Insertion sort takes $O(n^2)$ time. Rahmani sort algorithm gives much better performance than Insertion sort by quickly finding the position of the new element in the sorted sub array. In the following sub section, the salient features of Rahmani sort are being described.

### 3.2 The Abstract Procedure

The algorithm is in-place sorting algorithm. The original input array is logically partitioned into two sub arrays: a left sub array and a right sub array. In the beginning, the left sub array is consisting of only the first element of the original array and the right sub array is consisting of the remaining elements of the original array. The left sub array is sorted because it is having only a single element and by definition of sorting a single element is always sorted. The procedure of Rahmani sort for arranging the input array in ascending order is being describes as below:

### *ABSTRACT-PROCEDURE (A, n)*

1. The logical left sub array is sorted with the first element of the original array.
2. Find the actual position of the next element of the array in the left sub array by using the procedure iSearch*h*().
3. By finding the actual position of element in the sorted left sub array, iSearch*h*() returns the index where the next element is to be inserted.
4. The new element is inserted at that index.

The abstract form of the *Rahmani* sort algorithm can be implemented in two different ways. First method is based on an iterative procedure and the second one is based on recursive procedure. In this paper, only the first method is described which takes running time of O($n$log$n$).

### 3.3 The Algorithms

*Rahmani* sort algorithm is comprising of one sub procedure *iSearch()* with one main procedure *RahmaniSort()*. In the main algorithm *RahmaniSort()*, the element would be inserted in its proper position in the sorted left sub array after shifting all the elements of the sorted left sub array from that position to the right by one place till the last element. The shifting of elements starts from the right hand side to create a vacant position at the correctly found location by the sub procedures *iSearch()*. The *iSearch()* algorithm is designed to find the position of the largest element which is less than the key element stored in variable 'key'. After finding this position, each element of the left sub array which are in the right side of this position will be shifted to the right by one place. The shifting operation will start from the extreme right position to save each value from getting over-written. **$TILL$ $HERE$** We present our pseudo code for insertion sort as a procedure called INSERTIONSORT, which takes as a parameter an array : : n containing a sequence of length n that is to be sorted. (In the code, the number n of elements in A is denoted by A:length.) The algorithm sorts the input numbers in place: it rearranges the numbers within the array A, with at most a constant number of them stored outside the array at any time. The input array A contains the sorted output sequence when the INSERTION-SORT procedure is finished [Coreman].

Description of the variables are given below:

a $\Rightarrow$ Input array of items to be sorted.

n $\Rightarrow$ Number of elements in the array 'a'.

lower $\Rightarrow$ Lower index of the array 'a'.

upper $\Rightarrow$ Upper index of the array 'a'.

mid $\Rightarrow$ Middle index of the array 'a'.

### 3.3.1. Algorithm for Rahmani Sort

RAHMANI-SORT(a, $n$)

1. **for** $i \leftarrow 2$ to $n$ **do**
2.    **if** a$[i] \geq$ a$[i-1]$ **then**
3.       **continue**
4.    $key \leftarrow$ a$[i]$
5.    **if** a$[i] \leq$ a$[1]$ **then**
6.       $j \leftarrow 1$
7.    **else** $j \leftarrow$ ISEARCH(a, 1, $i-1$, $key$)
8.    $k \leftarrow i$
9.    **while** $k > j$ **do**
10.       a$[k] \leftarrow$ a$[k-1]$
11.       $k \leftarrow k-1$
12.    a$[j] \leftarrow key$

### 3.3.2. Algorithm for Iterative Search of the Location

ISEARCH(a, *lower*, *upper*, *key*)

1. **do**
2.    $mid \leftarrow (lower + upper)/2$

3.        **if** *key* = a[*mid*] **then**

4.           **return** *mid* + 1

5.        **if** *key* < a[*mid*] **then**

6.           *upper* ← *mid* – 1

7.        **else**

8.           *lower* ← *mid* + 1

9.    **while** *lower* ≤ *upper* **and** *key* != a[*mid*]

10. **return** *lower*

### 3.4 The Java Framework for Testing Performance of the Algorithms

A framework for comparing performance of sorting algorithms is a platform for implementing various sorting algorithm and works as an application to save time from repeatedly giving inputs to each algorithm and tabulating the outputs of each algorithm as records. The outputs of the algorithms are redirected to excel sheets in all three cases viz. best-case, worst-case, and average-case. This framework includes a class and many functions that can be used to process input, tabulate outputs. Designing a Java framework is useful that researchers can use to implement their algorithms/modules by writing codes very easily for their own algorithms. Frameworks saves from manual overhead of creating everything from scratch.

The framework to compare the performance of sorting algorithms is designed in Java for the simple reason of Java being the fastest among the triplet of <C++, Java, C\#> [13]. A simple but elegant and effective framework has been developed to test and compare the performance of sorting algorithms. Reason for using Java is its nature of strong variable typing, robust support by the VM for million size of data, capability of measuring CPU running time up to *nano* seconds *ns*.

A single class is used to develop the program in which many methods/code snippets are used, description of them is given below:

### 3.4.1 Data Sets Preparation

An input sequence given to an algorithm is called an *instance* of the sorting problem [3]. An *instance* of a problem may consist of some constraints. For the simplicity and ease of testing the correctness of the algorithms and comparing their performance with other well-known sorting algorithms, the *instance* in this paper is kept as integer type of data. The *instances* are picked up from the data sets prepared for all three different cases of the algorithms' analysis. For the average case, the data set is prepared by randomly generating integer values. The randomly created numbers are sorted in increasing order to give the best-case data set. Finally, the best-case data set is reversed to create the worst-case data set.

The size of integer data type in Java for 64-bit computing is 4 bytes. Data sets of various sizes: (500, 2500, 5000, 50000, 100000, 625000, 1250000, and 2500000) are randomly generated calling the *nextInt()* method of a random object of Random class and then kept in an array of exact size. The randomly created data set is taken as the average case data set and kept in the array, *averageArr*.

The two more data sets are prepared for the best case and worst case and kept in the array *bestArr* and *worstArr* respectively. For preparing the best-case data set, the original data set is sorted in ascending order and for the worst-case data set the best-case data set is reversed.

The following code snippet is a sample to have the idea of data sets preparation.

Random random = new Random();

int averageArr[] = new int[SIZE];

int bestArr[] = new int[SIZE];

int worstArr[] = new int[SIZE];

int RANGE = 2147483647; // maximum range of int value

```
// averageArr: for the average case
for (int i = 0; i < SIZE; i++)
{
        averageArr[i] = random.nextInt(RANGE);
        bestArr[i] = averageArr[i];
}
// bestArr: for the best case
int n = bestArr.length;
for (int i = 1; i < n; i++) // A simple sorting in increasing order
{
        boolean flag = true;
        // Bubble the largest up
        for(int j = 0; j < n - i; j++)
        {
                if(bestArr[j] > bestArr[j+1])
                {
                        flag = false;
                        // swap them
                        int val = bestArr[j];
                        bestArr[j] = bestArr[j+1];
                        bestArr[j+1] = val;
                }
        }
        if(flag)
        {
                break;
        }
}
// worstArr: for the worst case
for (int i = 0; i < bestArr.length; i++) // copy in reverse order
{
        worstArr[i] = bestArr[n-i-1];
}
```

The specific data sets prepared at one occasion are used for each algorithm's CPU running time measurement to truly compare their running time performances. After successfully measuring the running time of each algorithm with the current data set, another data set of larger size is similarly prepared and used for performance measurement of each algorithm.

### 3.4.2 Measuring CPU Running Times

Java provides many necessary static methods in the *System* class. For recording the system time two methods are available; *currentTimeMillis()* method returns the current time in *milli* seconds and *nanoTime()* method returns the current time in *nano* seconds. The later method is used for this purpose as it facilitates the most precise available system timer. The *nanoTime()* method is called once before the execution of the algorithm and its returned time is kept in a *long* type of variable *startTime* and it is once again called after the finish of execution of the algorithm and its returned time is kept into another *long* type of variable *stopTime*. So, the difference of the values of the two variables is the actual *execution* time of the algorithm excluding its input time and output time. So, the code snippet is as below:

averageStartTime = System.nanoTime();

ref.rahmaniSort(averageArr);

averageStopTime = System.nanoTime();

averageElapsedTime = averageStopTime - averageStartTime;

### 3.4.3 Recording the CPU running times

For recording the CPU running times (in the order of *nana* seconds) various Excel sheets are created. These Excel sheets record the running times of each algorithm for each data sets and for each case. For creating Excel workbook and sheets etc., the open-source *poi* library from the Apache Software Foundation is used. Please, see Figure

## 4 Analysis of Rahmani Sort

### 4.1 Disadvantage with Insertion sort

The Insertion-sort() is having some disadvantages. Firstly, it searches the location of the next element in the sorted left sub array sequentially. It is a failure to exploit the ordering of the elements in the left sub array. Secondly it is not able to differentiate between the average case scenario and the worst-case scenario. Therefore, the insertion sort algorithm is not able to attain the level of performance which it could have attained if these shortcomings would have been resolved.

### 4.2 Analysis of Insertion sort

For analysis of the algorithms, generalized mathematical expressions are derived with the cost functions in terms of input data size and finally expressed in the form of asymptotic notations: $\Omega$ for best case, $\Theta$ for average case, and O for worst case. The computational time complexity for individual steps is provided in **Tab. 2** that is based on the algorithm provided in Insertion-sort().

**Table 2:** Analysis of Insertion sort

| Insertion-sort() | Cost | General Formulae | Best Case | Average Case | Worst Case |
|---|---|---|---|---|---|
| **for** $j$ = 2 to $n$ **do** | $C_1$ | $n$ | $n$ | $n$ | $n$ |
| $key$ = a[$j$] | $C_2$ | $n-1$ | $n-1$ | $n-1$ | $n-1$ |
| $i = j$ - 1 | $C_3$ | $n-1$ | $n-1$ | $n-1$ | $n-1$ |
| **while** $i > 0$ **and** $key$ < a[$i$] **do** | $C_4$ | $\sum_{i=2}^{n} T_i$ | 1 | $\sum_{i=2}^{n}(\frac{i}{2}+1)$ | $\sum_{i=2}^{n} i$ |
| a[$i$ + 1] = a[$i$] | $C_5$ | $\sum_{i=2}^{n} T_i - 1$ | 0 | $\sum_{i=2}^{n} \frac{i}{2}$ | $\sum_{i=2}^{n}(i-1)$ |
| $i = i$ - 1 | $C_6$ | $\sum_{i=2}^{n} T_i - 1$ | 0 | $\sum_{i=2}^{n} \frac{i}{2}$ | $\sum_{i=2}^{n}(i-1)$ |
| a[$i$ + 1] = $key$ | $C_7$ | $n-1$ | $n-1$ | $n-1$ | $n-1$ |

### 4.2.1 Best Case Analysis

The best case of input instance for the sorting algorithms occurs when the elements are already in sorted order. Therefore, $T(n)$ for the best case can be derived as below:

$$T(n) = C_1 + C_2 \left[ \frac{(n+1)(n+2)}{2} - 3 \right] + C_3 \left[ \frac{n(n+1)}{2} - 1 \right]$$

$$= \frac{C_2 + C_3}{2} n^2 + \left[ C_1 + \frac{3C_2 + C_3}{2} \right] n + (-2C_2 - C_3)$$

$$T(n) = An^2 + Bn + C \tag{1}$$

where, $A = \frac{C_2 + C_3}{2}$, $B = C_1 + \frac{3C_2 + C_3}{2}$, and $C = -2C_2 - C_3$

Since Eq. 1 is a quadratic equation, time complexity of Insertion sort in best-case is O($n^2$).

The Eq. 1 is not correct.

### 4.2.2 Worst Case Analysis

The worst case of input instance for the sorting algorithms occurs when the elements are arranged in reverse order.

Therefore, *T(n)* for the worst case can be derived as below:

Since the Eq. 2 is a quadratic equation, algorithm's time complexity in average case is O($n^2$).

### 4.2.3 Average Case Analysis

The average case of input instance for the sorting algorithms occurs when the elements are in random order. This is the most expected case.

Therefore, *T(n)* for the average case can be derived as below:

Since the Eq. 3 is a quadratic equation, algorithm's time complexity in average case is $\Theta(n^2)$.

### 4.3 Rahmani Sort

The proposed sorting algorithm has improved the sorting performance by using an improved search mechanism to quickly find the place where the next element is to be inserted in a proper position of the sorted left sub array. This can be proved by showing the asymptotic notations in average-case as $\Theta$, best-case as $\Omega$, and worst-case as O for input data of size *n*. For obtaining the asymptotic notations in various cases, the algorithm's running steps are tabulated in an easily deductible form as shown in Tab. 3 for the iterative version of the algorithm. In the first column of the table, the algorithm's steps are mentioned with proper indentation for the ease of quick acquaintance with each step. In the second column of the table, the symbols $C_1$, $C_2$, …, $C_{12}$ represent the time (cost) incurred for executing each respective step by the CPU of the PC. In the third column, general formulae for the number of repetitions of the corresponding steps are written. Fourth, fifth and sixth columns are used to represent the number of repetitions of the steps in the best-case, average-case, and worst-case respectively. The computation model is assumed to be the well-known sequential RAM model having only one CPU with no concurrency.

**Table 3:** Analysis of Rahmani sort (iterative version)

| Rahmani-sort() | Cost | General Formulae | Best Case | Average Case | Worst Case |
|---|---|---|---|---|---|
| **for** $i \leftarrow 2$ to $n$ **do** | $C_1$ | $n$ | $n$ | $n$ | $n$ |
| **if** a$[i] \geq$ a$[i-1]$ **then** | $C_2$ | $n-1$ | $n-1$ | $n-1$ | $n-1$ |
| **continue** | $C_3$ | $n-1$ | $n-1$ | $\dfrac{n-1}{3}$ | $0$ |
| $key \leftarrow$ a$[i]$ | $C_4$ | $n-1$ | $0$ | $\dfrac{2(n-1)}{3}$ | $n-1$ |
| **if** a$[i] \leq$ a$[1]$ **then** | $C_5$ | $n-1$ | $0$ | $n-1$ | $n-1$ |
| $j \leftarrow 1$ | $C_6$ | $n-1$ | $0$ | $\dfrac{n-1}{3}$ | $n-1$ |
| **else** $j \leftarrow$ ISEARCH(a, 1, $i-1$, $key$) | $C_7$ | $\sum\limits_{i=2}^{n} T_i$ | $0$ | $\sum\limits_{i=2}^{n} \log i$ | $0$ |
| $k \leftarrow i$ | $C_8$ | $n-1$ | $0$ | $\dfrac{2(n-1)}{3}$ | $n-1$ |
| **while** $k > j$ **do** | $C_9$ | $\sum\limits_{i=2}^{n} T'_i$ | $0$ | $\sum\limits_{i=2}^{n} (\dfrac{i}{2}+1)$ | $\sum\limits_{i=2}^{n} i$ |
| a$[k] \leftarrow$ a$[k-1]$ | $C_{10}$ | $\sum\limits_{i=2}^{n} T'_i - 1$ | $0$ | $\sum\limits_{i=2}^{n} \dfrac{i}{2}$ | $\sum\limits_{i=2}^{n} (i-1)$ |
| $k \leftarrow k-1$ | $C_{11}$ | $\sum\limits_{i=2}^{n} T'_i - 1$ | $0$ | $\sum\limits_{i=2}^{n} \dfrac{i}{2}$ | $\sum\limits_{i=2}^{n} (i-1)$ |
| a$[j] \leftarrow key$ | $C_{12}$ | $n-1$ | $0$ | $\dfrac{2(n-1)}{3}$ | $n-1$ |

### 4.4 Proof of the Analysis

To prove the correctness of the analysis of the algorithm, the general expression of the cost function in terms of the input data size are derived in the form of asymptotic notations for best-case $\Omega$, average-case $\Theta$ and worst-case O are derived. For proving this, analysis of the algorithm is described in the following sub sections by taking all three cases into consideration.

### 4.4.1 Best case

The best-case scenario for the algorithm arises when the given array is already sorted in the same order in which the algorithm is asked to sort the input array. The algorithm checks this condition in step 2 through an *if-then* structure if a$[i] \geq$ a$[i-1]$. If this is the case then the element a$[i]$ is already in its proper position and hence no further work is expected from the algorithm for this element i.e., no any comparison or shifting is performed. The *continue* statement of step *3* inside the *if* structure, ensures that all the remaining statements in the current iteration of the for loop is ignored. Hence, *ISEARCH*() is never called in the algorithm. So, the time complexity of *ISEARCH*() does not put any effect in the overall time complexity of the algorithm. Therefore, all the entries in the column after step 3 is 0. In fact, all elements in the input array are in their proper position. So, the algorithm does the minimum possible work in this case.

Since, only first 3 steps are contributing to the overall time complexity of the algorithm, the total time taken by the algorithm is computed by taking the sum of the products of the corresponding time(cost) and repetition:

Since the Eq. 1 is a linear equation, time complexity of the algorithm in best-case is $\Omega(n)$.

### 4.4.2 Worst case

The worst case of the input instance for the sorting algorithms occurs when the elements are arranged in the reverse order i.e., the input array is sorted in descending order and the algorithm is asked to sort elements of the input array in ascending order or vice-versa. The *for* loop in step 1 is tested $n$ times with $n – 1$ time for executing its body and 1 more time when its condition is not satisfied. The *if* structure in step 2 is never satisfied and so *continue* statement inside the *if* structure is never executed. Since every next element is smaller (or probably equal) than the elements in the sorted left sub array, its proper position in the sorted array should be in its initial position (i.e., $i = 1$). This is ensured by the *if* structure of step 5 in the algorithm (*if* a[$i$] $\leq$ a[0]) which gives the value of $j$ as 1 (i.e., $j = 1$). The *else* clause of the step 7 in the *if-else* structure in the steps 5-7, is never executed in this case. Hence, *ISEARCH*() is never called in the algorithm. So, the time complexity of *ISEARCH*() does not put any effect in the overall time complexity of the algorithm. So, $T_i = 0$ and hence the entry in the column of step 7 is 0.

The while loop in step *7* is used for the purpose of shifting all the elements of the sorted left sub array right by one position starting from the right-most element of the sorted sub array so that the initial position of the array is vacant for the key to be inserted there. Therefore, $T_i' = i$ for this case. Hence the total time for execution is computed by taking only the contributing terms into consideration:

The Eq. 2 is a quadratic equation so its time complexity is O($n^2$). But it is much faster than other sorting algorithms. The nature of the algorithm is quadratic because of the steps 9, 10, and 11. The while loop in step 9 is implemented for the purpose of shifting the elements which are larger than the value of the key element. The algorithm can be made even faster if some alternate data structures are used wherein insertion operation is available as a primitive operation or as an API with efficiently implemented method for the insertion operation. In that case, $T_i' = 1$, in step 9. So, $T_i' - 1 = 0$ in steps 10 and 11. Therefore, the new expression for $T(n)$ will be as below:

So, its time complexity will become O($n$).

### 4.4.3 Average case

The average case of input instance for the sorting algorithms occurs when the elements are arranged in random order i.e., the elements in the input array may be partially sorted but the elements will neither be arranged in sorted order nor in reverse order. This case is the most expected and critical one where elements are collected in the input array in their natural order of arrival. The algorithm uses another sub algorithm, *ISEARCH*() for searching the position of the next element in the sorted left sub array. Moreover, since the control structure of the *ISEARCH*() is the same as the well-known binary search algorithm, it can safely be accepted that the average-time complexity of *ISEARCH*() is $\Theta(\log n)$. So, as shown in the Tab. 3, the value of $T_i = \log n$. Further, the next element picked up (i.e., the key) can be either larger than the right-most element of the sorted left sub array, or smaller than the left-most element in the sorted left sub array or larger than the left-most element and smaller than the right-most element of the sorted left sub array. It is assumed that all the three outcomes are equally likely. So, step 3 (*continue*), and step 6 ($j = 0$) will be executed $\frac{n-1}{3}$ times. Step 4 (*key* = a[$i$]), step 8 ($k = i$), and step 12 (a[$j$] = *key*) will be executed $\frac{2(n-1)}{3}$ times. In steps 9 to 11, selection of the value of $j$ is very crucial for the proper analysis of the algorithm in this case. It can be safely assumed that the proper position of the key is in the middle of the sorted left sub array. So, $j = \frac{i}{2}$. So, in step 9, $i - j + 1 = \frac{i}{2} + 1$ and in step 10 and 11, $i - j = \frac{i}{2}$. Therefore, $T(n)$ for the average case can be derived as below:

Since the Eq. 3 has a prominent $n^2$ term as the highest order term, by neglecting all the lower order terms and ignoring the coefficient of the highest order term, it is shown that asymptotic notation of the algorithm's time complexity in average case is $\Theta(n^2)$. In average case, the algorithm will perform slowly than both best case and worst case. It is slower than the worst case because it takes more time for searching the proper position of the key in the sorted left sub array apart from the shifting of elements to make a vacant position for the key to be inserted in the proper position of the sorted left sub array.

The Eq. 3 is a quadratic equation so its time complexity is $\Omega(n^2)$. But it is much faster than other sorting algorithms. The nature of the algorithm is quadratic because of the steps 9, 10, and 11. The while loop in step 9 is implemented for the purpose of shifting the elements which are larger than the value of the key element. The algorithm can be made faster if some alternate data structures like in case of worst case are used wherein insertion operation is available as a primitive operation or as an API with efficiently implemented method for the insertion operation. In that case, $T_i' = 1$, in step 9. So, $T_i' - 1 = 0$, in steps 10 and 11. Therefore, the new expression for $T(n)$ will be as below:

So, its time complexity will become $\Omega(n)$.

## 5 Results and Discussions

In this section, I have created a dataset containing random, ordered, and reverse ordered data as input data to the algorithms. Then I have described the experimental setup before discussing the results.

For the experimental set-up a program in core Java has been written. The program creates separate data sets for each of the three cases. The sizes of the data sets are taken as 500, 2500, 5000, 50000, 100000, 625000, 1250000, and 2500000 respectively. First of all, an array of the specific size is created then the array is populated with the random numbers generated by the program. This array is used as the input array for the average-case. These array elements are copied into another array. Then a sorting algorithm is applied on this array to sort the elements in the ascending order so that the array can be used as the input array for the best-case. Now the elements of the second array (best-case array) are copied into a third array in reverse order to be used as the input array for the worst-case. The outputs of the algorithms (in $nano$ seconds) is redirected to excel sheets. Then the whole process is repeated for the data sets of the next size. The experiment was repeatedly performed as many as ten times for recording the performance of the algorithms for varying time duration. In the following sub section, all these results are being tabulated with the exact CPU time required for the execution.

Comprehensive results for illustrating the performance comparison of Rahmani sort algorithm with other sorting algorithms are being given.

### 5.1 Dataset

Using a perfect dataset is one of the important considerations for establishing an experimental setup for measuring performance analysis of computer algorithms. In this paper, we have used datasets of randomly created positive integer values generated by a dedicated Java program, which are stored in text format. The datasets are of 3 categories of sizes 500, 2500, 5000, 50000, 100000, 625000, 1250000, and 2500000.

#### 5.1.1 Random sets

Sets of the specified sizes were created in random order to analyze average-case scenario.

#### 5.1.2 Sorted sets

The Random sets were first sorted and then used for the analysis of best-case scenario.

### 5.1.3 Reverse Sorted sets

The Sorted sets were reversed and then used for the analysis of worst-case scenario.

We have designed a Java program that used the Random class along with its *nextInt*() method. This class was instantiated for calling the *nextInt*() method. The value of the argument *range* in the *nextInt*() method was given *2147483647* to maintain as much uniqueness of elements in the data set as possible.

Besides, we have created another similar dataset with approximately 50\% sorted data items to provide evidence of dynamic footprint detection capability for which the results are provided in section \ref{dynamic-footprint}.

As, it can be seen from the above result sets (for the worst-case) that for the data size less than 8000, the performance of Merge sort is better than that of Rahmani sort while for the larger data size i.e., from 8000 to 64000, Rahmani sort beats Merge sort in performance (except a few exceptions). Performance of Rahmani sort as compared to Merge Sort is shown in the seventh column of result sets; where the Yes denoted that Rahmani sort is better while the No signifies that Merge sort excels.

In the above tables, the result sets of average cases for various sorting algorithms show that there is a mixed outcome for the performance of Rahmani sort as compared to Merge sort. For some set of data Rahmani sort is better while for others Merge sort is better. But, in maximum cases for the higher size of data the performance of Rahmani sort is significantly better than that of Merge sort.

Now, the following table provides the execution time comparison for the best-case scenario of various sorting algorithms.

### 5.2 Experimental Setup

The steps of the proposed algorithm have been written in a formal way suitable for implementing in a programming language. I have implemented the algorithms in Java because it is one of the most efficient one in terms of level of accuracy in CPU time (measured in nano seconds) and memory requirement [22-23]. I have used IntelliJ IDEA 2020.3.1 Edition (Runtime version 11.0.9.1) with JDK 1.8.0_231 for implementing the algorithms, testing the results and validating the research objectives [24-25].

The experiments are performed on Laptop PC HP EliteBook 840 G1, running with Processor Intel® Core™ i7-4600U CPU @ 2.10 GHz, 2701 Mhz, 2 Core(s), 4 Logical Processor(s) with 8 GB RAM, and 64-bit OS (Windows 10 Pro) platform.

### 5.3 Graphical Comparison

Fig. 2 gives a graphical comparison of different sorting algorithms in worst case by using line chart. Here, lower the graph slop better the performance and vice-versa.

In the above graph Rahmani Sort shows some remarkably better performance as compared to Merge Sort. Fig. 3 gives a graphical comparison of different sorting algorithms in average case by using line chart.

Fig. 4 gives a graphical comparison of different sorting algorithms in best case by using line chart. This graph shows that in best case, Rahmani Sort cannot defeat Merge sort.

## 6 Conclusion

In this research paper a new sort algorithm namely Rahmani Sort algorithm has been proposed. The algorithm has been developed by using an improved binary search-like mechanism to quickly determine the sorted position of the new element into the sorted left portion of the array as compared to the classical insertion sort algorithm. The improved binary search-like mechanism has been adapted from the original binary search algorithm for applying it in the proposed sorting algorithm. The promising results have been obtained through rigorous experimentation for computing the running times of the proposed and other sorting algorithms using the elegance and power of Java programming language which facilitates the getting the computation times to the level of nanoseconds. Although the computation times of all the cases can be computed, I have shown the results of average case computational time of my algorithm and other important existing algorithms like classical insertion sort algorithm, bubble sort, merge sort and quick sort algorithms. The results are very promising. The average-case computation time of my algorithm is found to be far better than the computation times of bubble sort and classical insertion sort algorithms. Its average-case time complexity is similar as merge sort algorithm. Only quick sort is better than my algorithm in the average-case. But my algorithm is beating the quick sort in the worst-case. The performance of the new algorithm can be safeguarded for the best-case, by adapting to an early verification of the sortedness of the elements and hence avoiding the unnecessary delay in the improved binary search mechanism employed in the algorithm. In the future, the algorithm can be further improved by designing a good methodology for finding the position of the elements even more quickly and also avoiding the shift operations required to make the necessary vacant position for the insertion of the current element. To escape from the more time-consuming shift operations, a linked list data structure can be used. At present, I am using arrays. but it is almost very difficult task to apply improved binary search in linked list. In future, main focus would be on how to build a better way of achieving this goal. One possible way to solve the issue is avoiding using linked list and using an array list and applying the Java library function sort() of the Array class.

The reason that why my algorithm is not performing well in the average-case is that some extra effort is required in first searching the place of insertion where the next item would be placed and then shifting all the elements towards the right of this location in the sorted sub-array. This introduces an additional amount of work to be done. So, the algorithm takes more time to sort the elements in the average-case.

**Funding Statement:** The author received no specific funding for this study.